\documentclass[iop]{emulateapj}

\usepackage{graphicx}  
\usepackage{dcolumn}   
\usepackage{bm}        
\usepackage{amssymb}   
\usepackage{hyperref}
\usepackage{amsmath}
\usepackage{color}
\usepackage{multirow}

\usepackage{color}

\DeclareMathOperator{\sech}{sech}

\begin{document}

\newcommand{\va}{v_{\textsc a}}
\newcommand{\taua}{\tau_{\textsc a}}
\newcommand{\gammai}{\gamma_i}
\newcommand{\ai}{a_i}

\graphicspath{{graphs/}}

\title{Magnetic Reconnection: Recursive current sheet collapse triggered by ``ideal'' tearing }

\author{Anna Tenerani} \author{Marco Velli}  
\affiliation{EPSS, UCLA, Los Angeles, CA}
\email{annatenerani@epss.ucla.edu}
\author{Antonio Franco Rappazzo}\affiliation{Advanced Heliophysics, Pasadena, CA}
\author{Fulvia Pucci}
\affiliation{Universit\`a degli studi di Roma Tor Vergata, Rome, Italy}

\begin{abstract}
We study,  by means of MHD  simulations, the onset and evolution of fast reconnection via the ``ideal" tearing mode  within a collapsing current sheet at high Lundquist numbers ($S\gg10^4$). We first confirm that as the collapse proceeds, fast reconnection is triggered well before a Sweet-Parker type configuration can form: during the linear stage plasmoids rapidly grow in a few Alfv\'en times when the predicted ``ideal" tearing threshold $S^{-1/3}$ is approached from above; after the linear phase of the initial instability, X-points collapse and reform nonlinearly. We show that these give rise to a hierarchy of tearing events repeating faster and faster on current sheets at ever smaller scales,  corresponding to the triggering of ``ideal" tearing at the renormalized Lundquist number. In resistive MHD this process should end with the formation of sub-critical ($S \leq10^4$) Sweet Parker sheets at microscopic scales.  We present a simple model describing the nonlinear recursive evolution which explains the timescale of the disruption of the initial sheet. 
\end{abstract}


\maketitle

\section{Introduction}

Magnetic reconnection is thought to provide the pathway for energy release in solar flares and other phenomena where energy is accumulated in the magnetic fields and currents of ionized high temperature plasmas. What has remained difficult to understand has been the triggering and speed of the process itself, which, given the extremely large Lundquist ($S$) and Reynolds ($R$) numbers, implies that currents must collapse to extremely small scales before anything close to realistic timescales are approached.

Indeed, an active region in the solar corona has a typical spatial scale $L$ of about {$L\simeq10^{9}$}cm, a magnetic field  $B\simeq 50$G, density $\rho\simeq 10^9$cm$^{-3}$ and a temperature  $T\simeq 10^6$K, hence a macroscopic Lundquist number $S\simeq10^{13}$. A flare emits a total energy of about $10^{32}$ergs on a typical timescale of a few minutes, which cannot be explained by simple magnetic field diffusion or by a reconnecting instability occurring on the macroscopic scale itself. It has been suggested then that the reconnection
trigger might only be provided by kinetic effects, beyond the (resistive) magnetohydrodynamic description of the plasma,  or, alternatively, by the fast plasmoid instability of Sweet-Parker current sheets (\cite{cassak_2013} and references therein).

It has been predicted recently,  using linear stability analyses, that the tearing mode instability should grow on ideal timescales (``ideal'' tearing, or IT)  once the inverse aspect ratio $a/L$ of a current sheet reaches a scale $a/L\sim S^{-1/3}$, preventing the formation of the paradigmatic Sweet-Parker current sheet (SP) for which $a_{\text{\sc sp}}/L\sim S^{-1/2}$, about 150 times smaller than  $S^{-1/3}$ for $S\simeq 10^{13}$ and even smaller for greater $S$. Even more interestingly, the (linear) asymptotic tearing instability in thin current sheets at arbitrary aspect ratios predicts that the  maximum growth rate normalized to the macroscopic Alfv\'en time  increases nonlinearly with the aspect ratio, $\gamma_m\taua\sim(L/a)^{3/2}$ (see also eq.~(\ref{km}) below): this suggests that collapse to the critical threshold thickness may provide the trigger for fast reconnection, where ``fast'' here is meant for dynamics occurring on the ideal timescale~\citep{pucci_2014, tenerani_2015,landi_2015,delsarto}.  

At the intermediate Lundquist numbers usual to simulations, say, $S\simeq 10^4-10^5$, the presence of plasma flows into (inflows), and along (outflows) the current sheet affects the scaling of the critical threshold for IT by inducing the formation of more elongated current sheets, that is, of layers having inverse aspect ratios $a/L\sim S^{-\alpha_c}$, with  $\alpha_c>1/3$~\citep{velli_unp}.  Therefore, with the Lundquist number not sufficiently large, the distinction between the IT scaling and the SP scaling might seem academic. Nevertheless, the IT current sheet instability, as typical for the tearing mode, produces a quasi-singular inner layer $\delta_i$, whose inverse aspect ratio follows the SP scaling $\delta_i/L\sim S^{-1/2}$. Such inner layer is however non-stationary, with inflows and outflows increasing exponentially in time---their ratio incidentally does not follow the SP scaling, but rather $u_{in}/u_{out}\sim S^{-1/3}$---together with the amplitude of the reconnecting field: seen from this inner layer,  one may interpret the dynamics in terms of the ``embedded reconnection" scenario~\citep{cassak_2009}, leading to greater energy storage and more efficient dissipation at the extremely large $S$ values of the solar corona.

Here we study the onset and evolution of the tearing instability within a single collapsing current sheet by means of resistive MHD simulations at $S=10^6$. { In the following paragraphs we first  show that the transition to a fast tearing mode instability takes place  during the collapse when the predicted threshold inverse aspect ratio $a/L\sim S^{-1/3}$ is reached. Secondly, we show that the secondary current sheets formed nonlinearly give rise to recursive tearing instabilities at increasingly smaller scales and faster super-Alfv\'enic timescales, which may also be well described by the flow-modified ``ideal" tearing criterion and instability. The nonlinear evolution leads to a complete disruption in a timescale estimated at 0.05 percent of the macroscopic  Alfv\'en time once initial tearing is triggered, which we model here in a corrected version of the fractal reconnection scenario first proposed by~\citep{shibata}. Though the two-dimensional instability of current sheets and its subsequent nonlinear evolution within resistive MHD has been studied before, it is shown here for the first time that the IT instability scenario, appropriately modified by the effect of the reconnection velocity flows, provides a quantitative description of the various stages in the evolution of a reconnecting current sheet, including hierarchical secondary island formation and disruption. Our results also provide a coherent framework through which previous simulations~\citep{loureiro_2005,lapenta,daughton_PRL_2009, battac_2009} may be more completely understood. }

\section{Trigger of fast reconnection: scenario}
\label{theo}
Let us consider a (local) current sheet  of inverse aspect ratio $a/L$,  as the one given in eq.~(\ref{har}), with Alfv\'en speed $\va$ and magnetic diffusivity $\eta$, in a high $S=L\va/\eta\gg1$, constant density plasma. We assume, as in~\citep{uzdensky_2014}, that the current sheet is collapsing via some external process on a timescale~$\tau_c\sim a\, (da/dt)^{-1}$, which we take of the order of the ideal timescale, $\tau_c\gtrsim\taua= L/\va$. We extend the linear analysis of the  incompressible tearing instability to this time-dependent case by defining an instantaneous growth rate $\gamma(k,t)$, $k$ being the wave number along the sheet, where explicit time-dependence is introduced by the evolving aspect ratio $L/a$ itself.
 
In the static case, two regimes describe the unstable spectrum of a current sheet: the so called small $\Delta^{\prime}$ regime, for wavelengths close to the instability threshold  (or const.-$\psi$~\citep{FKR}, for $ka\lesssim 1$ for our equilibrium), and the large $\Delta^{\prime}$ regime  (or non const.-$\psi$~\citep{coppi_1976}, for $ka\ll1$). These two regimes  have wave vectors $k$ that lie to the right and to the left of the fastest growing mode  $k_m$,  $k_m<k<1/a$ and $0<k<k_m$, respectively~\citep{battac_2009, loureiro_2013,delsarto}. The wave vector and the growth rate of the fastest growing mode are given by~\citep{pucci_2014}:   
\begin{equation}
k_ma\sim S^{-1/4}(L/a)^{1/4}, \quad \gamma_m\taua\sim S^{-1/2}\left(L/a\right)^{3/2}.
\label{km}
\end{equation}
Similarly to eqs.~(\ref{km}), the growth rates in the small and large $\Delta^{\prime}$ regimes also increase with the aspect ratio; however, their values tend to zero in the asymptotic limit  $S\rightarrow\infty$, and the full $\gamma(k)$ dispersion relation becomes rapidly peaked around~$k_m$ (cfr.  Fig.~1 of~\citep{pucci_2014}). Therefore, both the small and large $\Delta^{\prime}$ regimes can be neglected in the framework of fast reconnection, since the development of the instability is controlled by the evolution of the fastest growing mode described by eqs.~(\ref{km}). Take the  current sheet given by eq.~(\ref{har}), for example: the unstable spectrum lies within the range $2\pi/L\lesssim k<1/a$. From this condition and the first of eqs.~(\ref{km}) it follows that the regime described by eqs.~(\ref{km}) always exists once $k_mL\gtrsim 2\pi$, i.e. once $a/L\lesssim 0.2\,S^{-1/5}$.  
The second of eqs.~(\ref{km}) shows that the growth rate becomes ideal when approaching the critical width $a/L\sim S^{-1/3}$  (much smaller than $S^{-1/5}$) from above: the transition to ``ideal'' tearing during collapse therefore occurs on the fastest growing mode.
In conclusion, a collapse with $\tau_c\gtrsim \taua$ naturally drives a sudden switch from a quasi-stable state---growth rate depending on  a negative power of $S$---to an ideally unstable~one. Otherwise, the disruption of the current sheet at widths thicker than critical (as considered in~\citep{uzdensky_2014}) would require an infinite time for $S\rightarrow\infty$. 

We now consider the example of an exponential collapse. {Exponentially thinning current sheets are observed in simulations of solar and  stellar coronal heating \citep{rappazzo}, in 3D MHD turbulence~\citep{brachet_2013,grauer_2000} of interest to a broad range of astrophysical phenomena, as well as  in the nonlinear stage of the classic tearing instability itself~\citep{ali_2014}. }
\section{Numerical set up}
We consider  a plasma with homogeneous density $\rho_0$ and pressure $p_0$. The background magnetic field ${\bf B}_0$ describes an exponentially shrinking current sheet and is given by 
\begin{equation}
{\bf B_0}= B_0\tanh\left[  {y}/{a(t)} \right]{\bf\hat x}+ B_0\sech \left[ {y}/{a(t)} \right]\bf\hat z,
\label{har}
\end{equation}
where the half width $a(t)$ is prescribed and parametrized in time by   
\begin{equation}
a(t)= a_0\exp^{-t/\tau_c}+a_\infty  (1 -\exp^{-t /\tau_c}).
\label{coll}
\end{equation}
We employ a  2.5D compressible MHD code periodic in~$x$ and with non-reflecting boundary conditions along the inhomogeneous~$y$ direction~\citep{landi_2005}. We assume an adiabatic closure, with index $\Gamma=5/3$, a scalar resistivity, and a Newtonian viscous stress tensor with Prandtl number $P=1$. With this choice, viscosity does not affect the scalings of IT significantly~\citep{tenerani_2015}. The collapse of the background magnetic field is obtained by adding a { source} term $\mathcal{F}$  of the form
\begin{equation}
 \bm{\mathcal{ F}}=\left(\frac{1}{a}\frac{\partial a}{\partial t}\right)\,{y}\,\frac{\partial {\bf B}_0}{\partial y}
\end{equation}
in Faraday's equation. Magnetic and velocity fields are normalized to $B_0$ and to $\va=B_0/\sqrt{4\pi \rho_0}$, respectively,  lengths to the macroscopic length $L$, time to $\taua$, and density and pressure  to  $\rho_0$ and  $B_0^2/4\pi$, respectively. It is useful to introduce also the normalized flux function $\psi$, such that ${\bf B}=\boldsymbol\nabla\times\psi{\bf \hat z}$. We set $p_0=0.8$ and $S=10^6$. Instability is seeded with a random noise of small amplitude. Because of the wide range of scales to be resolved in the $y$ direction, we employ an inhomogeneous grid with increasing resolution at the neutral line. The simulation box has normalized dimensions $L_x\times L_y=2\pi\times 0.96$ (runs~1--3) and $L_x\times L_y=2\pi\times 0.46$ (run~4), with $2048\times1024$ mesh points.  {Resolution at the neutral line is $\Delta y=0.0001$, which allows to resolve the diffusion region of the IT, that scales as $\delta_i/L\sim S^{-1/2}$~\citep{pucci_2014}}. Since for $a\simeq L$  the instability is extremely slow, we start from $a_0=0.1\,L$. This is a good compromise for limiting computational time while retaining a sufficient dynamical range for the collapse. 
 \begin{table}
 \caption{\label{runs}Simulation parameters: Lundquist number,  collapsing time, asymptotic half  width, onset time, nonlinear time, and half width reached at the end of the linear stage.}
\begin{ruledtabular}
\begin{tabular}{ccccccc}
run \#			&$S$&$\tau_c/\taua$&$a_\infty/L$&$\tau_*/\taua$&$\tau_{nl}/\taua$&	$a(\tau_{nl})/L$\\	
\hline
run 1 &\multicolumn{1}{c}{\multirow{4}{*}{$10^6$} } & 	1 &	$S^{-1/3}$	&3&16&0.01       \\
run 2 &&$4$	&$S^{-1/3}$	&11.5	&$22$&0.0105	\\
run 3 &&$10$	&$S^{-1/3}$	&16	&$36$&0.013	\\
run 4 &&$1$	&$S^{-1/2}$	&3	&4.5&0.0024	\\
\end{tabular}
\end{ruledtabular}
\end{table}
\section{Results}
\begin{figure}[b]
\begin{center}
\includegraphics[width=0.48 \textwidth]{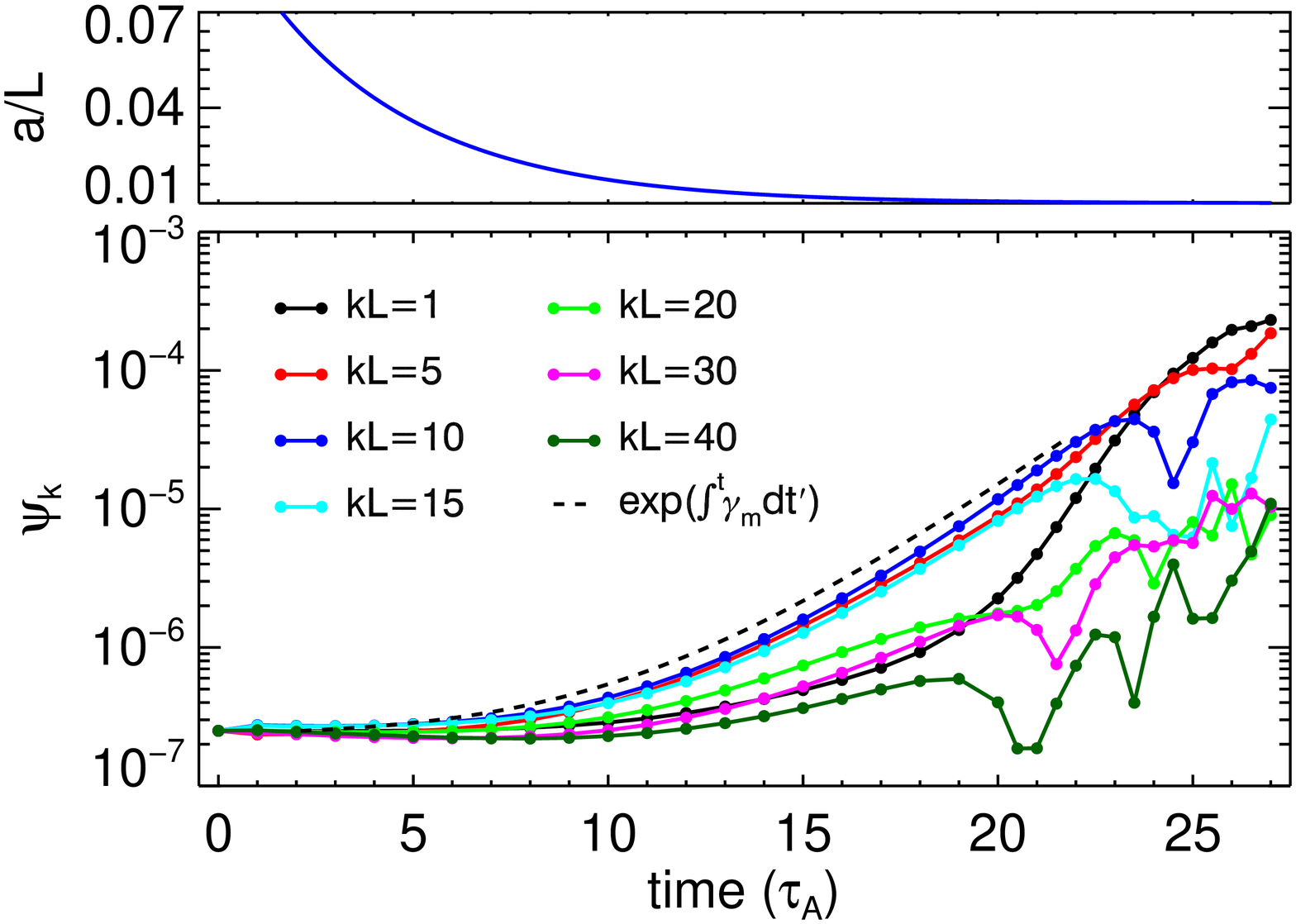}
\caption{Upper panel:  $a/L$ vs. time. Lower panel: temporal evolution of the amplitude of some unstable modes~$\psi_k(y)$ at the neutral line $y=0$ (run 2).}
\label{run1}
\end{center}
\end{figure}

 In Table~\ref{runs} we list the background and main dynamical parameters for each run, and in Fig.~\ref{run1} we show for reference  the temporal evolution of some unstable Fourier modes of the flux function $\psi_k(y)$, in $y=0$, from run~2. The growth rate and wave number of the fastest growing mode of the primary tearing at the critical IT threshold will be labelled $\gamma_i$ and $k_i$, respectively. In particular, in our simulations we find $k_iL=10$ and $\gamma_i\taua=0.46$, in agreement with theory~\citep{tenerani_2015}.
\subsection{Linear stage}\label{linear}
Tearing onset takes place at time $\tau_*$ { such that} $\gamma_m(\tau_*)\tau_c=1$, where $\gamma_m=0.46\,S^{-1/2}(a/L)^{-3/2}$. After onset, the most unstable  modes grow according to the WKB solution $\psi_m(t)\sim\exp[\int^t \gamma_m(t^\prime)dt^\prime]$,  represented by the dashed line in Fig.~\ref{run1}. In runs~1--3 we consider different cases of collapse in which $a_\infty/L=S^{-1/3}$. In all these runs the linear stage is ultimately dominated by modes close to the ideally unstable one  $k_iL= 10$.  Run~4 forces a collapse of the initial sheet down to $a_\infty/L=S^{-1/2}$, but proves that inverse aspect ratios $a/L\ll S^{-1/3}$ cannot be formed: the tearing mode indeed ignites right after $t=2.4\,\taua$, at which $a/L=S^{-1/3}$, and a large number of islands rapidly pop up. Even at the end of the linear stage {(see Section~\ref{earlyNL})} the collapsing sheet thickness is twice the corresponding SP thickness at the given $S$ and $P$~\citep{tenerani_2015}.    
\subsection{Early nonlinear stage}
\label{earlyNL}
Nonlinearity becomes important at~$\tau_{nl}$, when the  width~$w$ of magnetic islands is of the order of the width of the inner diffusion layer~$\delta$. The fastest growing mode has $\delta_m$ which scales with the aspect ratio as $\delta_m/a\sim (a\va/\eta)^{-1/4}$~\citep{loureiro_2013},  hence $\delta_m/a\sim S^{-1/4}(L/a)^{1/4}$. By using the definition~$w/a\simeq2\,\sqrt{\psi_m/a}$~\citep{bisk_book}, we estimate the amplitude  $\psi_m(\tau_{nl})\simeq{0.25} \,S^{-1/2} \left[a(\tau_{nl})/L\right]^{1/2}\simeq{2.5}\times10^{-5}$, in agreement with simulations. Nonlinear effects lead to the competition of two  processes~\citep{malara_1991}: on the one hand, islands start to merge through an inverse cascade process from the fastest growing modes to smaller wave numbers, see, e.g., the inverse cascade from $k_iL=10$ towards $kL=5$ and below  starting at $\tau_{nl}=22\,\taua$ (Fig.~\ref{run1}); on the other hand, X-point collapse, and subsequent  secondary current sheet formation, takes place during the further nonlinear growth of the islands, as is typical for strongly unstable modes far from the small $\Delta^{\prime}$ regime~\citep{jemella}, in lieu of the slow, algebraic Rutherford growth~\citep{Ruther}. 
\subsection{Nonlinear { stage: recursive X-point collapse}}
\begin{figure}[htbp]
\begin{center}
\includegraphics[width=0.48 \textwidth]{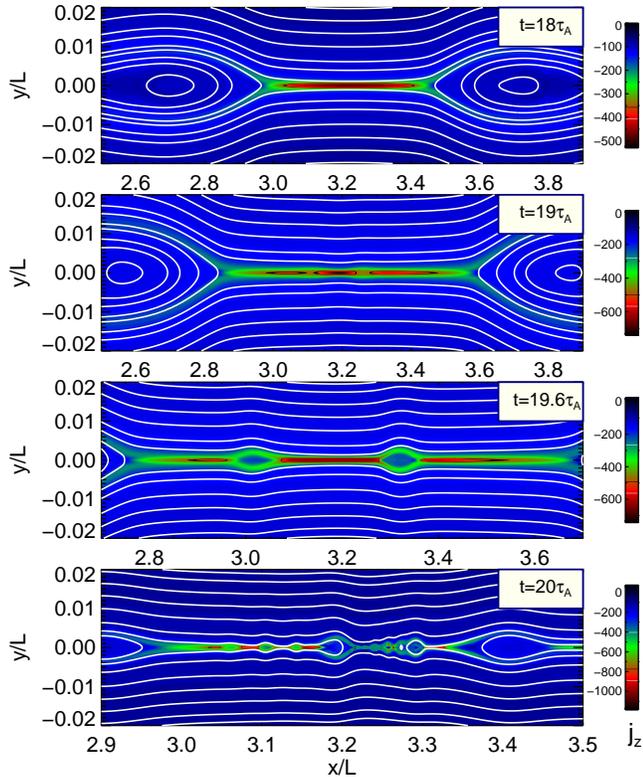}
\caption{Blow-up of the { recursive} ``ideal'' tearing {instability} in run~1: contour plot of the out-of-plane current density $j_z$ (color coded) and magnetic field lines (white).  {Notice that the whole domain is much larger, $0\leq x/L<2\pi$ and $-0.48\leq y/L\leq0.48$.}}
\label{phi_run0}
\end{center}
\end{figure}

\begin{figure}[b]
\begin{center}
\includegraphics[width=0.48 \textwidth]{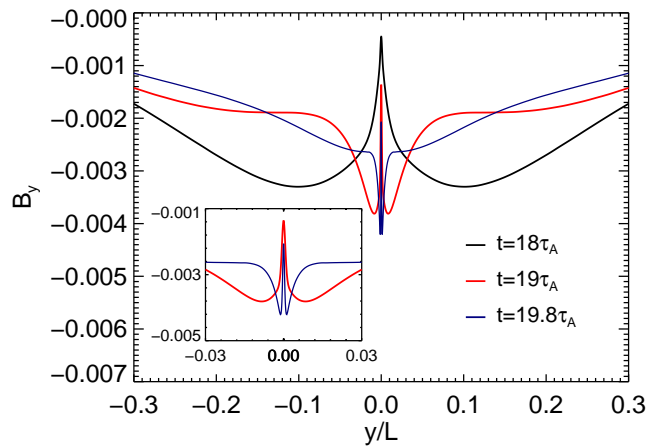}
\caption{{Profile of} $B_y(y,x_0)$ in $x_0/L=3.2$ at $t=18\,\taua$ (primary tearing, $n=0$), $t=19\,\taua$ ($n=1$),  and  $t=19.8\,\taua$ ($n=2$). { The inset  is a blow-up of $B_{y}(y,x_0)$ at $t=19\,\taua$ and $t=19.8\,\taua$.}}
\label{eigen}
\end{center}
\end{figure}

\begin{figure}[t]
\begin{center}
\includegraphics[width=0.48 \textwidth]{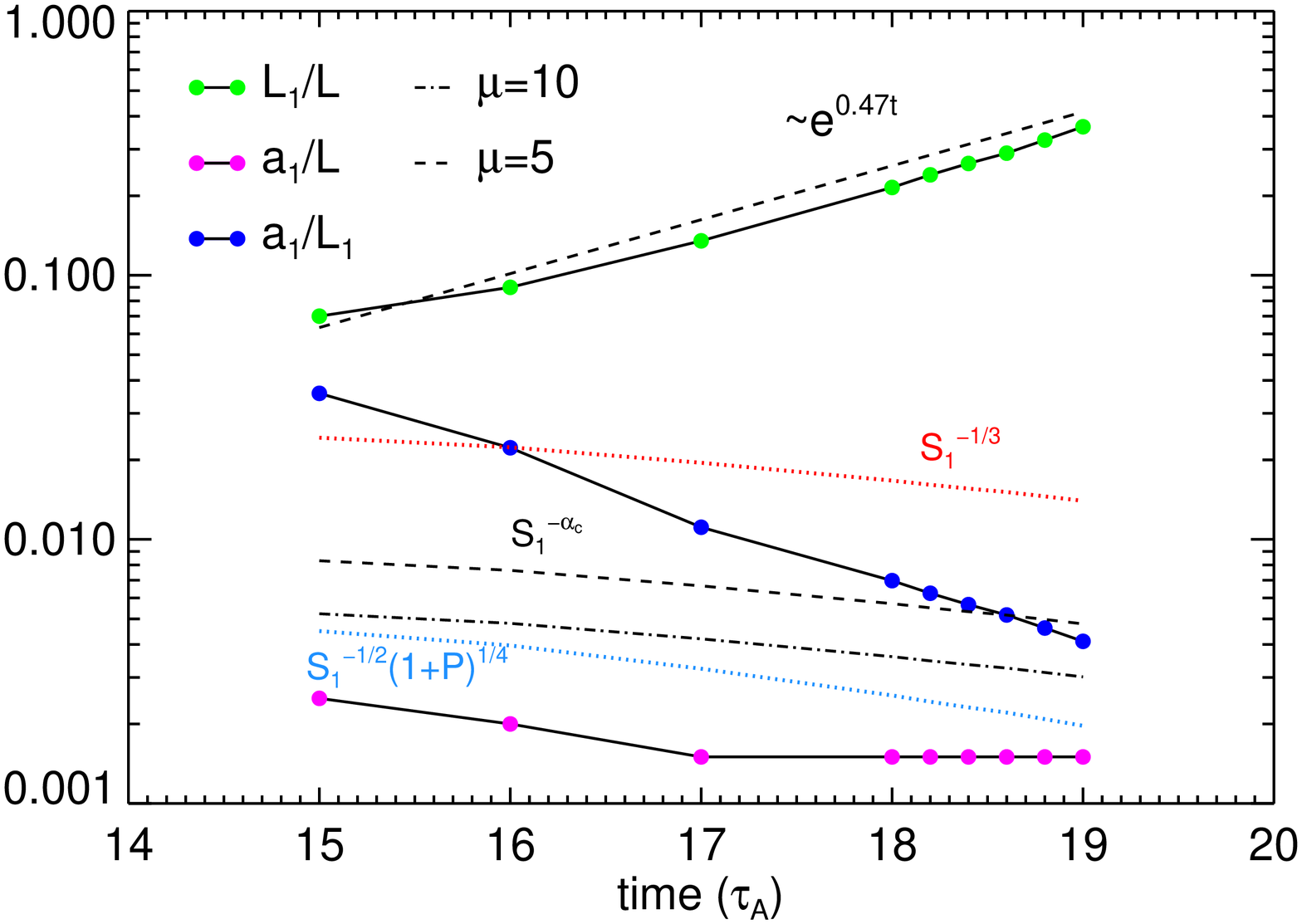}
\caption{Length $L_1/L$ (green), width $a_1/L$ (magenta),} and inverse aspect ratio $a_1/L_1$ (blue) of the first secondary current sheet vs. time (run~1). The red dotted line represents the critical threshold $S^{-1/3}$, and the light blue   dotted line the viscous $a_{\text{\sc sp}}/L$. Black dotted and dash-dotted lines represent the critical threshold with outflow effects, given by~eq.~(\ref{outf}).
\label{current}
\end{center}
\end{figure}

We analyze the fully nonlinear stage using run~1 as a reference, shown in Fig.~\ref{phi_run0}.
X-point collapse proceeds leading to the formation of secondary plasmoids (magnetic islands)~\citep{loureiro_2005,lapenta} and, next, a recursive process of super-Alfv\'enic secondary layer formation and disruption, takes place~{\citep{shibata,daughton_PRL_2009,battac_2009}}. Fig.~\ref{phi_run0} shows a temporal sequence of recursive plasmoid formation, taking place at the center, near the flow stagnation point of the original instability. Previously, such recursive reconnection has been modeled as a succession of unstable SP layers~{\citep{loureiro_2005, daughton_PRL_2009, cassak_2009, huang_2010, uzdensky_2010, Loureiro_2012}}. Our simulations show instead that it is driven by onset of the IT mode, triggered by the dynamical \emph{lengthening} of sheets to the local critical threshold, in a way similar to that discussed in section~\ref{linear}. As the recursive X-point collapse occurs in steps, we label the half length of the $n$-th current sheet $L_n$, the inverse aspect ratio $a_n/L_n$, the normalized (half) width of the inner diffusion layer $\delta_{n}/L_n$, the local Lundquist number $S_n=(L_n/L)\,S$ and  Alfv\'en time $\tau_{{\text{\sc a}},n}=(L_n/L)\,\taua$. Referring to Fig.~\ref{phi_run0},  the {recursive X-point collapse} starts with the formation of $L_1$ from the primary diffusion region of thickness $\delta_i$, hence $a_1\simeq\delta_i$. Current sheet $L_1$ becomes unstable, less than $3\,\taua$ after the end of the linear stage (second panel), consistent with what was found in~\citep{landi_2015}. Here we show that IT triggering within $L_1$ induces the growth of two plasmoids and of a second current  $L_2$, with  $a_2\simeq\delta_1$ (third panel), which is itself destroyed by multiple plasmoids  in about $0.5\,\taua$ (fourth panel). In Fig.~\ref{eigen} we plot  $B_y(y,x_0)$, in $x_0/L=3.2$, at three different times.  The shown profiles of the magnetic field closely resemble the tearing mode eigenfunction, and display a clear hierarchical structure  within each of the inner diffusion layers.    
To further analyze the tearing onset  inside such secondary layers, in Fig.~\ref{current} we  display  $L_1/L$ (in green), {$a_1/L$ (magenta)}, and $a_1/L_1$ (in blue) as a function of  time, from $t=15\,\taua$, up to $t=19\,\taua$. During the collapse process the width remains almost constant, $a_1\simeq\delta_i\sim LS^{-1/2}$ (here $\delta_i\simeq0.0015$), while the length $L_1$ stretches at about a rate $\dot L_1/L_1\simeq 1/\tau_{c,1}$, $\tau_{c,1}\simeq2\taua$. Fig.~\ref{current} shows that $a_1/L_1$ crosses the critical threshold $S_1^{-1/3}$ (red dotted line) at $t\simeq16\,\taua$. At that time $L_1\simeq0.1\,L$, thus $S_1\simeq10^{{5}}$,  $P=1$, and  linear theory predicts $\gamma_{i,1}\tau_{\text{\sc a},1}\simeq0.47$, or $\gamma_{i,1}\taua\simeq2.3\gg\tau_{c,1}^{-1}\taua$. In the light of the previous discussion, we would expect  $L_1$ to disrupt more rapidly than in $\simeq3\,\taua$, though on the same order of magnitude timescale. On the other hand, there is a stabilizing effect from outflows forming along $L_1$~\citep{bulanov}: these may induce the formation of thinner sheets having inverse aspect ratio $a_n/L_n\sim S_n^{-\alpha_c}$, with $\alpha_c>1/3$. Based on the empirical observation that flows stabilize SP up to $S_c\simeq10^4$, one can  extend the IT theory to find a critical $\alpha_c$ which includes the effect of inflow-outflows~\citep{velli_unp}:  
\begin{equation}
\alpha_c=\frac{2\log \mu+\log S_n}{3\log S_n}.
\label{outf}
\end{equation}
In eq.~(\ref{outf}),  $\mu=\Gamma_v/(f\gamma_i)$ is the ratio of the plasmoid evacuation rate  to the maximum growth rate, corrected by a threshold factor $f\simeq0.5-0.1$. The value $\mu=10$~\citep{bisk_book} yields the observed $\alpha_c=1/2$ for $S_n = 10^4$, while as expected  $\displaystyle\lim_{S_n\to\infty}\alpha_c=1/3$. Note that, as shown below, $S_n$ is a decreasing function of $n$. The  black dotted and dash-dotted lines in Fig.~\ref{current}, correspond to $S_1^{-\alpha_c}$ at two different $\mu$, while the light blue dotted line corresponds to $a_{\text{\sc sp,1}}/L_1$. Though our Lundquist numbers are not extremely large, $L_1$ still disrupts before reaching the SP width. The trend of the data plotted in Fig.~\ref{current} goes in the direction of our scenario, that is, that the fast tearing instability is triggered within collapsing current sheets once the critical threshold is reached. Taking into account the increased Alfv\'en speed  due to pile-up just outside the inner diffusion region as considered by~\citep{cassak_2009} does not change the trend of our results.    Interestingly, the discussion of flux pile-up and embedded reconnection led~\citep{cassak_2009} to independently find a similar scaling $\sim S^{-1/3}$ for the inverse aspect ratio of unstable current sheets starting from an initial secondary Sweet-Parker sheet. Their analysis starts from a different  line of thought, in the search for the proper criterion to destabilize an embedded SP sheet, and proceeds via a linearization of the fields in the neighborhood of the neutral line. It then finds a scaling with $S$ for   the aspect ratio of the embedded layer scaling in a way similar to  IT.  On the other hand, the IT scaling derives from requiring an $S$-independent growth rate from the complete eigenmode analysis of tearing instability theory, and finds that the entire sheet inverse aspect ratio scales as $S^{-1/3}$. The results therefore have different origin, and for the IT case the presence of flows modifies the scaling exponent $\alpha_c$ according to eq.~(\ref{outf}): incidentally, the latter expression is in excellent agreement with the aspect ratios at which plasmoids are observed to be ejected in the~\citep{cassak_2009}  paper.

\section{Discussion}
We have shown that a collapsing current sheet disrupts due to the triggering of fast reconnection as predicted by the IT model. This process is shown to proceed { recursively,} giving rise to a hierarchy of fast tearing instabilities at ever smaller scales. Let us  model the { recursive} collapse of X-points, in the spirit of the original fractal reconnection model, with the following differences:  first, we don't consider SP as the initial condition; second, we use the width $a_n$, derived from the IT scalings, to find $L_n$. Neglecting for simplicity viscous and outflow effects, imposing that  $a_n/L_{n-1}\sim S_{n-1}^{-1/2}$ (i.e., $a_n\simeq\delta_{n-1}$), and that IT onset occurs at $a_n/L_n\sim S_n^{-1/3}$, we find:
\begin{equation}
L_n=L\,S^{-1+(3/4)^n},\,\, \tau_{\text{\sc a},n}=L_n\taua,\,\, S_n=S^{(3/4)^n}.
\label{power}
\end{equation}
For $n\rightarrow\infty$ then  $\tau_{\text{\sc a},n}\rightarrow\taua\,S^{-1}$, and, as expected, $S_n\rightarrow1$ and $a_n/L_n\rightarrow1$. However, after a number $n_*$ of steps $S_{n_*}\simeq 10^4$, at which we expect to reach a stable SP. For $S=10^{13}$ (solar corona)  $n_*\simeq4$, and the length $L_n$ suddenly drops to microscopic scales, $L_1/L\simeq6\times10^{-4}$, $L_2/L\simeq2\times10^{-6}$, $L_3/L\simeq3\times10^{-8}$, $L_4/L\simeq10^{-9}$. The {recursive X-point collapse} after the first trigger thus lasts a time interval $\tau_{tot}\simeq\sum_{n=1}^{n_*}\tau_{\text{\sc a},n}\simeq5\times10^{-4}\taua$, one order of magnitude less than the upper limit given by~\citep{shibata}. Higher resolution simulations are nevertheless necessary to assess how the { nonlinear evolution} saturates. Our scenario can be modified to include kinetic scales which might be reached dynamically in typical astrophysical systems~\citep{cassak_PRL_2005, daughton_PRL_2009}. In this case we expect a change of the power laws~(\ref{power}) as different scalings describe the physics at the critical threshold in the kinetic regime~\citep{delsarto}. 

\acknowledgements{We thank D. Del Sarto for interesting discussions and  useful comments. This work was supported by NASA via the Solar Probe Plus Observatory Scientist grant.}

\end{document}